\documentclass[aps,pra,twocolumn,groupedaddress,showpacs]{revtex4}

\usepackage{graphicx}    

\usepackage{psfrag}
\usepackage{amsmath}

\begin{document}

\title{Evidence of antiblockade in an ultracold Rydberg gas}

\author{Thomas Amthor}
\email[]{amthor@physi.uni-heidelberg.de}
\affiliation{Physikalisches Institut, Universit{\"a}t Heidelberg,
  Philosophenweg 12, 69120 Heidelberg, Germany}
\author{Christian Giese}
\altaffiliation{Permanent address: Universit{\"a}t Freiburg, Hermann-Herder-Str. 3, 79104 Freiburg, Germany}
\author{Christoph S. Hofmann}
\author{Matthias Weidem{\"u}ller}
\email[]{weidemueller@physi.uni-heidelberg.de}
\affiliation{Physikalisches Institut, Universit{\"a}t Heidelberg,
  Philosophenweg 12, 69120 Heidelberg, Germany}

\date{\today}

\begin{abstract}
We present the experimental observation of the antiblockade in an ultracold Rydberg gas
recently proposed by Ates et al. [Phys. Rev. Lett. 98, 023002 (2007)].
Our approach allows the control of the pair distribution in the gas and is based on a
strong coupling of one transition in an atomic three-level system, while introducing
specific detunings of the other transition.
When the coupling energy matches the interaction energy of the Rydberg long-range interactions,
the otherwise blocked excitation of close pairs becomes possible.
A time-resolved spectroscopic measurement of the Penning ionization signal is used to identify
slight variations in the Rydberg pair distribution of a random arrangement of atoms.
A model based on a pair interaction Hamiltonian is presented which well reproduces our experimental 
observations and allows one to analyze the distribution of nearest neighbor distances.
\end{abstract}


\maketitle

The long-range character of strong Rydberg--Rydberg interactions gives rise to a
variety of phenomena.
The interaction-induced suppression or blockade of excitation is one of the most striking effects.
It has been demonstrated in a number of different realizations using ultracold gases of interacting Rydberg atoms \cite{tong2004,singer2004,vogt2006,vogt2007,heidemann2007,urban2009,gaetan2009}.
One of the driving forces in the study of Rydberg blockade and spatial structure is the possible application for quantum information processing \cite{jaksch2000,lukin2001,brion2007}.
In certain configurations, the blockade can be overcome and atoms pairs can selectively be
excited at short distance.
Specific geometric arrangements of three atoms have been identified which reduce
a blocking due to dipole interactions \cite{pohl2009}. A so-called antiblockade has recently been
proposed for a three-level two-photon Rydberg excitation scheme
where the van der Waals (vdW) interaction energy at a given atomic distance corresponds to the
Autler-Townes splitting induced by the lower transition \cite{ates2007b}.
By exploiting the blockade and antiblockade in a controlled way, spatial correlations can be imposed
on the excited atoms in an otherwise randomly arranged ultracold gas.
Specific pair distances can thus be preferred or suppressed, which is of particular interest not only
for quantum information, but for a variety of applications of controlled interactions, such
as resonant energy transfer \cite{anderson1998,mourachko1998,westermann2006,ditzhuijzen2008},
the creation of long-range molecules \cite{greene2000,boisseau2002,bendkowsky2009},
or the study of quantum mechanical transport phenomena in chains \cite{cote2006,mulken2007}.
Some recent experiments use small optical dipole traps to fix the distance between Rydberg atoms
\cite{urban2009,gaetan2009}, but this technique does not allow to prepare a large number of
pairs at specific range of distances simultaneously.

\begin{figure}
 \includegraphics[scale=0.75]{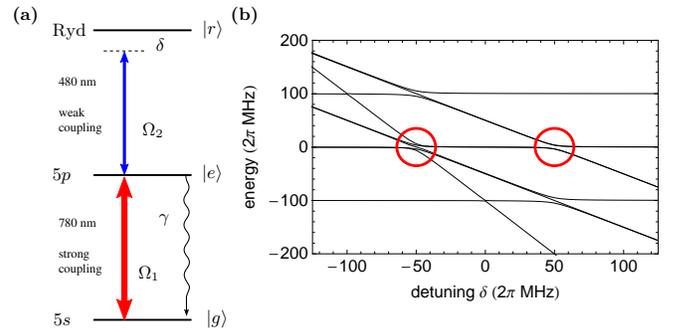}
 \caption{(a) Excitation level scheme. The states 5S and 5P of $^{87}$Rb are
	coupled by a strong laser field, while the weak upper transition can
	be brought to resonance with the Rydberg state 62D.
	(b) Dressed eigenstates of an interacting two-atom system for
	$\Omega_1=V_\mathrm{int}=2\pi\times 100\,$MHz. The circles indicate the detunings where the
	ground state couples to Rydberg states. Only at $\delta=-2\pi\times 50\,$MHz the
	state $|rr\rangle$ is populated.
  }
 \label{fig:ATscheme}
\end{figure}

In the characterization of the antiblockade \cite{ates2007b}, Ates et al. considered a lattice with fixed
lattice constant, while tuning the interaction strength between the Rydberg atoms by changing the
principal quantum number.
An increased two-photon Rydberg excitation probability in a three-level scheme (see
Fig.~\ref{fig:ATscheme}) was predicted when interaction energy and Rabi frequency $\Omega_1$ of
the lower transition match.
In our experimental demonstration of this effect, a complementary approach is followed. Starting
from an unstructured gas, we show that atomic pairs can be resonantly excited even at interatomic
separations below the blockade radius.
While keeping the quantum number fixed, the interaction strength is instead matched by picking 
specific atom pairs with appropriate interatomic distance.
Our method of measuring this modification of the pair distribution by the antiblockade is based on the dynamics of atomic pairs under the influence of the forces exerted by the interactions, as explored
in Ref. \cite{amthor2007}. Given an attractive vdW interaction, the collision times of close
pairs are shorter resulting in Penning ionization at shorter times.
In this way we are able to identify even slight variations in the nearest neighbor distance distribution.
The problem of antiblockade in a randomly distributed gas is briefly discussed in \cite{ates2007},
however we consider the proposed detection technique of ion counting statistics less practical
compared to the collision method used here.

\begin{figure}
 \includegraphics[scale=.55]{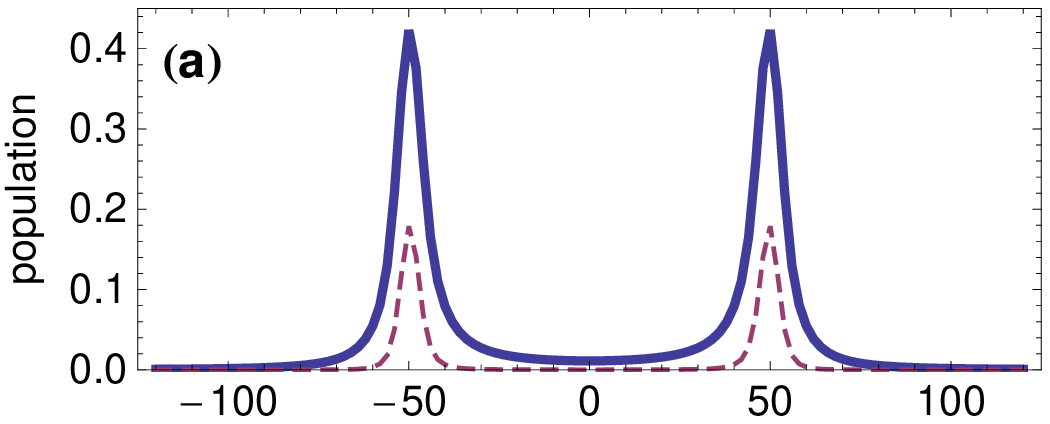}\vspace{-2mm}
 \includegraphics[scale=.55]{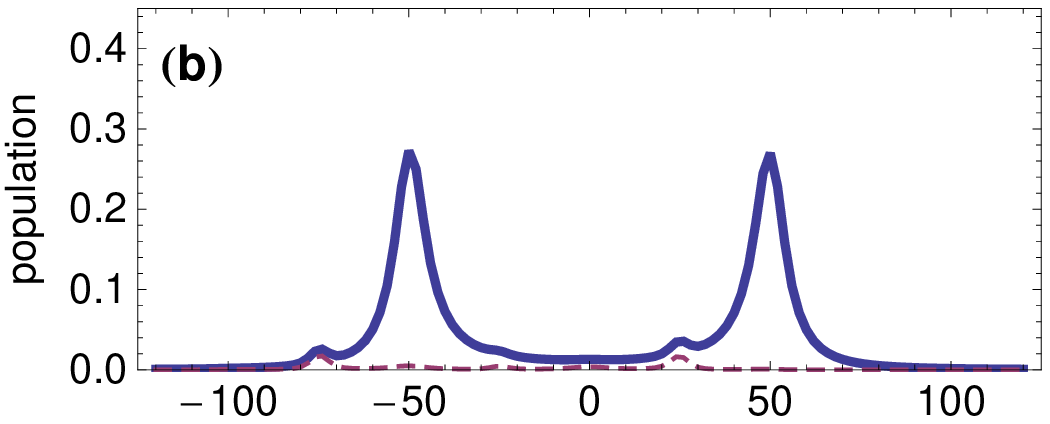}\vspace{-2mm}
 \includegraphics[scale=.55]{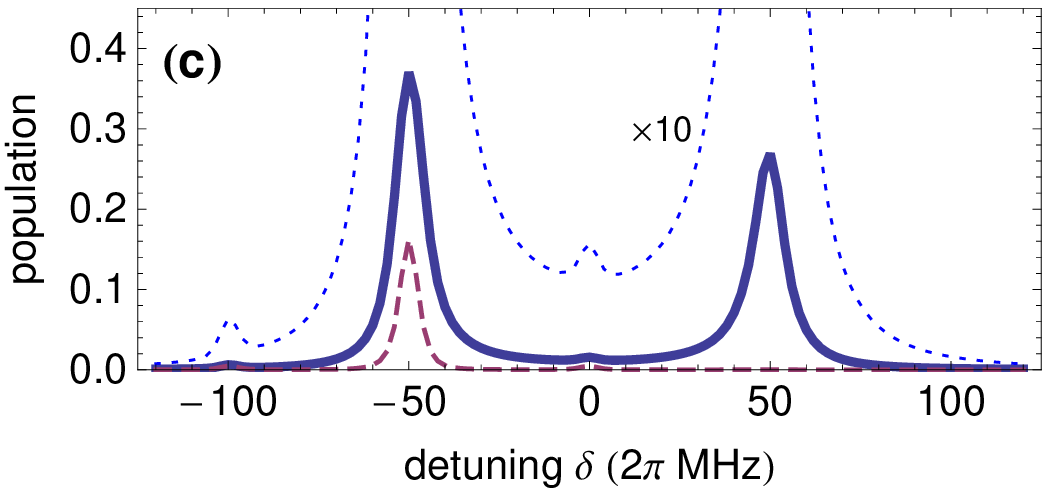}
 \caption{Calculated Rydberg population spectra for $\Omega_1=2\pi\times 100\,$MHz and
	fixed pair distances corresponding to interaction energies $V_\mathrm{int}$ of
	(a) 0\,MHz, (b) $2\pi\times 50\,$MHz, (c) $2\pi\times 100\,$MHz.
	The solid blue lines represent the
	total Rydberg population per atom, the dashed red lines are the population of the
	Rydberg pair state $|rr\rangle$. The dotted line in (c) is the total population scaled by
	a factor of 10. Parameters and laser pulse shape correspond to the experimental settings.}
 \label{fig:PairSpectra}
\end{figure}

We describe the system in terms of an interacting two-atom system with the three atomic levels
$|g\rangle$, $|e\rangle$, $|r\rangle$, corresponding to 5s, 5p, and Rydberg level, respectively.
This gives the total of nine two-body states, i.e. $|gg\rangle$, $|ge\rangle$, $|eg\rangle$, $|ee\rangle$, $|gr\rangle$, $|rg\rangle$, $|er\rangle$, $|re\rangle$, $|rr\rangle$.
The calculations are based on the Hamiltonian $H = H_\mathrm{at}+H_\mathrm{int}$, where $H_\mathrm{at}$
represents the couplings within the three-level systems and
$H_\mathrm{int}=V_\mathrm{int}(R)|rr\rangle\langle rr|$
is the vdW interaction with $V_\mathrm{int}(R)=-C_6^{\mathrm{eff}}/R^6$.
The interaction of two Rydberg atoms with non-vanishing angular momentum leads to a large number of
interaction potentials due to the many 
possible symmetries. We introduce an effective vdW coefficient $C_6^\mathrm{eff}$ to reduce the
calculation to a three-level system \cite{amthor2007}.
We solve the Master equation for the density matrix, $\dot\sigma=-\frac{i}{\hbar}[H,\sigma]+\Gamma$, where
$\Gamma$ contains the decay of the intermediate state $\gamma=2\pi\times 6\,$MHz as well as the laser
linewidths ($\sim 2\pi\times 1\,$MHz) as additional dephasing terms.
The spontaneous and black-body induced decay of the Rydberg level can be neglected
here because the time scale of observation is well below the lifetime of the Rydberg state chosen in the
experiment.

The dressed energy levels of two interacting three-level systems are depicted in
Fig.~\ref{fig:ATscheme}(b) as a function of the upper laser detuning $\delta$.
The slopes of the lines indicate the Rydberg character of the states: lines with slope $-1$ correspond to
levels containing one Rydberg excitation, lines with slope $-2$ correspond to a doubly excited Rydberg
state. The circles indicate the situation where population can be transferred to
Rydberg states, resulting in the well-known Autler-Townes doublet if the upper transition is probed
by a weak probe beam.
If the splitting $\Omega_1$ is equal to the interaction energy $V_{\mathrm{int}}$,
the state $|gg\rangle$ is strongly coupled to the doubly excited
Rydberg state $|rr\rangle$ at $\delta=-\Omega_1/2$, but not at $\delta=+\Omega_1/2$.
This leads to efficient population of $|rr\rangle$ even for close pairs and high interaction energies. 

Calculated Rydberg excitation spectra for pairs at three different separations can be found in
Fig.~\ref{fig:PairSpectra}. 
Without interaction (atoms far apart), a symmetric Autler-Townes spectrum is visible
(Fig.~\ref{fig:PairSpectra}(a)).
An interaction energy of $\Omega_1/2$ causes a considerable blockade of excitation, with the
population of the state $|rr\rangle$ being completely suppressed at the two resonances but
slightly enhanced on the red wing of each excitation line (Fig.~\ref{fig:PairSpectra}(b)).
If the excitation energy is further increased to match the lower Rabi frequency $\Omega_1$,
the doubly excited state $|rr\rangle$ is again populated at $\delta=-\Omega_1/2$, but is still completely
suppressed at $\delta=+\Omega_1/2$ (Fig.~\ref{fig:PairSpectra}(c)).
Enhanced excitation is then also possible at $\delta=0\,$MHz, but is less efficient compared to
$\delta=-\Omega_1/2$ (see dotted curve).
The antiblockade discussion in \cite{ates2007b} is restricted to this peak at zero detuning.
Introducing a finite detuning of $\delta=-\Omega_1/2$, the antiblockade effect becomes
much more pronounced. In addition, it is now possible to directly compare a blocked regime
(positive detuning) with an anti-blocked regime (negative detuning) in a single spectrum.

\begin{figure*}
 \includegraphics[scale=0.9]{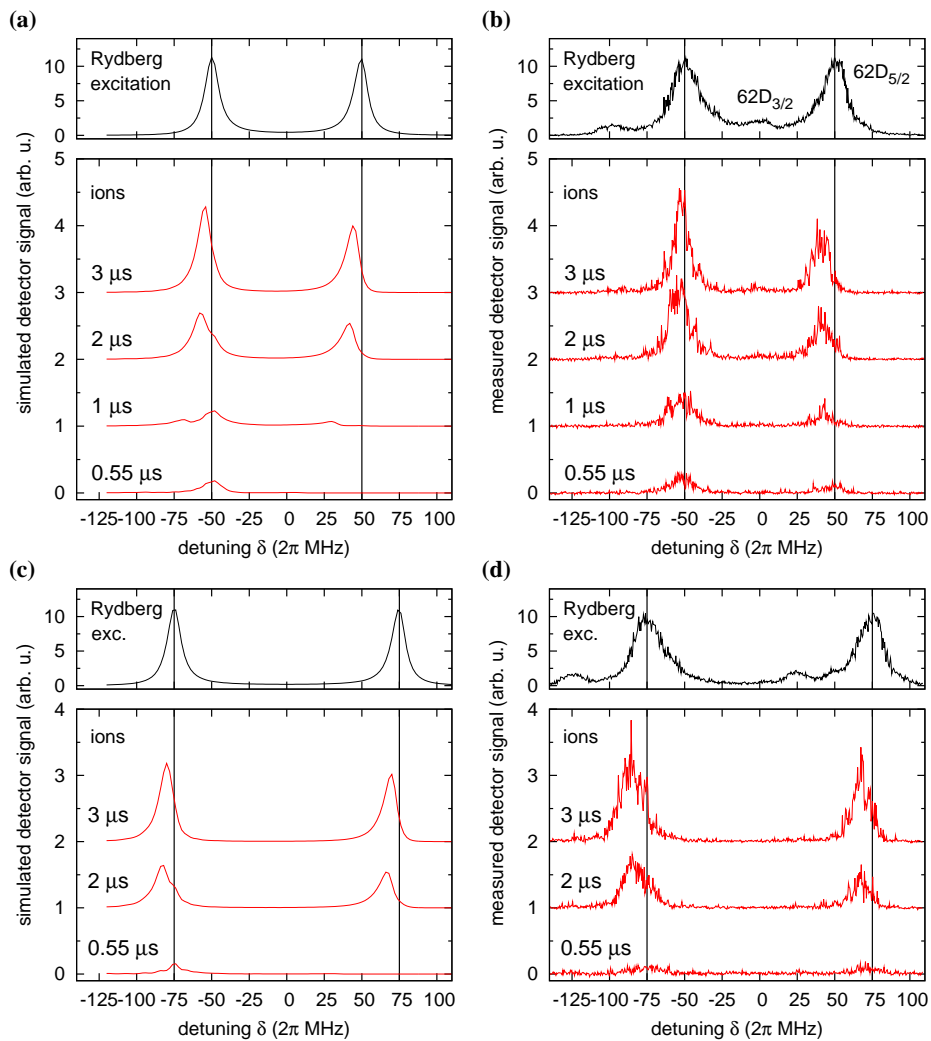}
 \caption{
	Comparison between calculated and measured 62D Rydberg excitation spectra and
	ionization spectra.
	Left column: Model calculations
	for (a) $\Omega_1=2\pi\times 100\,$MHz, and (c) $\Omega_1=2\pi\times 150\,$MHz,
	and a density of the trapped ground state atoms of $7\times 10^{9}\,$cm$^{-3}$.
	Right column: measurements for the same parameters, respectively.
	The upper graphs in each set (black) show the Rydberg population directly after
	excitation. Below (red) the ion signal at different delay times is shown, offset
	vertically for clarity. The respective delay $\Delta t$ is indicated at each trace.
	In both model and measurement, ions appear earlier on the red-detuned
	Autler-Townes component, and the asymmetry is more pronounced for smaller $\Omega_1$.
	A red-shift of the maximum ionization relative to each peak in the excitation spectrum is
	always clearly visible. Vertical lines have been drawn at $\pm$50\,MHz
	and $\pm$75\,MHz to emphasize this.
	}
 \label{fig:IonizationAttractiveAT62D} 
\end{figure*}

In order to predict the outcome of a measurement in an unstructured cloud,
one determines weighted averages of spectra at different interatomic spacings representing the actual
distribution of pair distances \cite{hertz1909}.
The results of such a calculation for our experimental parameters is presented in
Fig.~\ref{fig:IonizationAttractiveAT62D}(a,c) for two different values of $\Omega_1$.
The same scaling factor is used for all graphs to match the experimental detector signal of
the Rydberg excitation.
The upper graph in each set 
(black) represents the total Rydberg excitation with contributions of $|gr\rangle$, $|rg\rangle$,
$|er\rangle$, $|ge\rangle$, and $|rr\rangle$, averaged over all nearest-neighbor distances.
The lower graphs (red) show only the population of $|rr\rangle$ as an average of all pair
distances at which the atoms are able to collide within the specified time delay.
Assuming that each colliding pair
leads to Penning ionization, these spectra correspond directly to the expected ion signal.
For these calculations an attractive vdW interaction with $C_6^\mathrm{eff}=5\times 10^{20}\,$a.u.
is assumed.
This value is consistent with long-range potential calculations \cite{singer2005b} for the state
62D$_{5/2}$ which yield values of the same order of magnitude for all possible molecular symmetries.
The collision time depends mainly on the long-range
part of the initially present attractive interaction potentials.
The additional excitation of close pairs for negative detuning is not discernible in
the total Rydberg population. The number of potentially colliding pairs, however,
differs significantly for the two peaks, which is the signature of the antiblockade.

To experimentally demonstrate the effect, atoms confined in a magneto-optical
trap are excited to the 62D state with a 50\,ns pulse of the upper transition, while the lower
transition laser is
turned on with a fixed Rabi frequency of $2\pi\times 100\,$MHz or $2\pi\times 150\,$MHz.
While the lower (red) laser illuminates
the whole trapped atom cloud, the upper (blue) laser is focussed to a waist of 37\,$\mu$m, resulting
in a peak Rabi frequency of $2\pi\times 10.6\,$MHz in the center.
After a delay time $\Delta t$, an electric field ramp is applied to drive the ions onto a micro-channel
plate detector. Ions already present due to Penning ionization are detected first,
while the remaining Rydberg atoms are only ionized when
the field is strong enough and can thus be distinguished in the detector signal.
The experimental cycle is repeated every 70\,ms. Each cycle yields one data point in the
measured Rydberg and ionization spectra presented in Fig.~\ref{fig:IonizationAttractiveAT62D}(b,d).
The structure of the 62D$_{3/2}$ and 62D$_{5/2}$ lines appears twice in the spectrum, 
separated by the lower Rabi frequency. We restrict our discussion to 62D$_{5/2}$, as
62D$_{3/2}$ is only weakly excited. As expected from the above calculations and from the observations
in \cite{amthor2007}, ions appear first on the red-detuned wings of each of the 62D$_{5/2}$ lines.

In accordance with theory, there is an additional asymmetry between the two Autler-Townes components.
The ionization is stronger and starts earlier on the component which appears at lower frequencies,
as expected from the above reasoning.
Clearly, the excitation of close pair states is suppressed for the peak at positive detuning, while
it is allowed (anti-blocked) for the peak at negative detuning.
The distinctness of this asymmetry depends on the availability of atom pairs for which
the interaction potential corresponds to $\Omega_1$.
As the ground state pair density for a separation $R$ is proportional to $R^2$,
more pairs are available at larger distances,
which is why the asymmetry becomes more pronounced for smaller Rabi splittings, as can be seen both
in the calculation and in the measurement.
The underestimation of the peak widths in the simulation can be explained by the fact that we 
did not consider the spatial intensity profile of the upper excitation laser but instead only used
the Rabi frequency present in the center of the beam to minimize calculation time.
Averaging over the radial distribution of blue Rabi frequencies has been found to account for
a slight broadening of the excitation peaks \cite{deiglmayr2006}.

\begin{figure}
  \includegraphics[scale=.65]{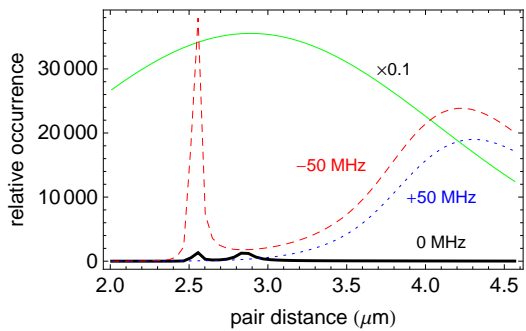}
 \caption{Calculated nearest-neighbor pair distribution of $|rr\rangle$ states
	for $\Omega_1=2\pi\times 100\,$MHz and $\delta=0$ (black solid),
	$\delta=2\pi\times 50\,$MHz (blue, dotted), and $\delta=-2\pi\times 50\,$MHz (red, dashed).
	The solid green curve above shows the ground state
	nearest neighbor distribution, scaled by a factor of $0.1$ for better visibility.}
 \label{fig:PairDistributionAT62Dmodel}
\end{figure}

The difference in the distribution of pair distances at the position of the two Autler-Townes
peaks $\delta=\pm\Omega_1/2$ is plotted in Fig.~\ref{fig:PairDistributionAT62Dmodel}.
The green solid line represents the distribution of nearest-neighbor distances in the cloud.
At $\delta=+\Omega_1/2$ (blue dotted line) the population of $|rr\rangle$ is strongly suppressed at
distances below 3.5\,$\mu$m.
At $\delta=-\Omega_1/2$ (red dashed line) additional pairs at small distances ($\sim 2\,\mu$m)
are present where the pair distance leads to an interaction energy close to $\Omega_1$.
At zero detuning (solid black line) there is also a finite excitation probability at small distances,
but the excitation is still comparatively small because of the strong Autler-Townes splitting. 

In conclusion, we have observed and modeled the effects of an antiblockade in an interacting Rydberg gas
excited with a strong coupling laser at the lower transition of a three-level system.
We used time-resolved ionization detection as a method to monitor the distribution of excited
pair distances which allowed us to clearly observe additionaly excited pairs at small distances
out of a large distribution.
The model calculation based on interacting atom pairs shows excellent agreement with the measured
ionization dynamics.
Making use of detuned excitation, we are able to directly manipulate the Rydberg pair distribution
through long-range interactions and to compare a blocked and an anti-blocked situation
in a single spectrum.
The asymmetry between the two ionization peaks is a direct probe for the strength of Rydberg interactions.
Our method may be used as an initialization or detection step in quantum gate experiments requiring
large numbers of atom pairs.
In future experiments it may become possible to observe the antiblockade effect in its most
drastic manifestation, when the atoms are arranged in a lattice with fixed interatomic distances.

The authors acknowledge financial support by the
Deutsche Forschungsgemeinschaft (grant no. WE2661/10-1).
We thank C. Ates, T. Pohl, and J.-M. Rost for stimulating discussions.

\bibliographystyle{apsrev}


\end{document}